\documentstyle[aps,eqsecnum,preprint,floats,epsf,epsfig]{revtex}
\begin{document}
\def\be{\begin{eqnarray}}
\def\en{\end{eqnarray}}
\def\non{\nonumber}
\def\la{\langle}
\def\ra{\rangle}
\def\nc{N_c^{\rm eff}}
\def\vp{\varepsilon}
\def\B{{\cal B}}
\def\up{\uparrow}
\def\dw{\downarrow}
\def\vma{{_{V-A}}}
\def\vpa{{_{V+A}}}
\def\smp{{_{S-P}}}
\def\spp{{_{S+P}}}
\def\J{{J/\psi}}
\def\ov{\overline}
\def\Lqcd{{\Lambda_{\rm QCD}}}
\def\pr{{\sl Phys. Rev.}~}
\def\prl{{\sl Phys. Rev. Lett.}~}
\def\pl{{\sl Phys. Lett.}~}
\def\np{{\sl Nucl. Phys.}~}
\def\zp{{\sl Z. Phys.}~}
\def\lsim{ {\ \lower-1.2pt\vbox{\hbox{\rlap{$<$}\lower5pt\vbox{\hbox{$\sim$}
}}}\ } }
\def\gsim{ {\ \lower-1.2pt\vbox{\hbox{\rlap{$>$}\lower5pt\vbox{\hbox{$\sim$}
}}}\ } }

\font\el=cmbx10 scaled \magstep2{\obeylines\hfill January, 2002}

\vskip 1.5 cm

\centerline{\large\bf Charmful Baryonic $B$ Decays $\bar B^0\to
\Lambda_c \,\bar p$ and $\ov B\to \Lambda_c \,\bar p\,\pi(\rho)$}
\bigskip
\centerline{\bf Hai-Yang Cheng$^{1,2,3}$ and Kwei-Chou Yang$^{4}$}
\medskip
\centerline{$^1$ Institute of Physics, Academia Sinica}
\centerline{Taipei, Taiwan 115, Republic of China}
\medskip
\centerline{$^2$ Physics Department, Brookhaven National
Laboratory} \centerline{Upton, New York 11973}
\medskip
\centerline{$^3$ C.N. Yang Institute for Theoretical Physics,
State University of New York} \centerline{Stony Brook, New York
11794}
\medskip
\centerline{$^4$ Department of Physics, Chung Yuan Christian
University} \centerline{Chung-Li, Taiwan 320, Republic of China}
\bigskip
\bigskip
\centerline{\bf Abstract}
\bigskip
{\small We study the two-body and three-body charmful baryonic $B$
decays: $\ov B^0\to \Lambda_c \,\bar p$ and $\ov B\to \Lambda_c
\,\bar p\,\pi(\rho)$. The factorizable $W$-exchange contribution
to $\ov B^0\to\Lambda_c\,\bar p$ is negligible. Applying the bag
model to evaluate the baryon-to-baryon weak transition matrix
element, we find $\B(\ov B^0\to\Lambda_c\,\bar p)\lsim 1.1\times
10^{-5}|g_{B^0p\Sigma_b^+}/ 6|^2$ with $g_{B^0p\Sigma_b^+}$ being
a strong coupling for the decay $\Sigma_b^+\to \ov B^0 p$ and
hence the predicted branching ratio is well below the current
experimental limit. The factorizable contributions to
$B^-\to\Lambda_c\,\bar p\pi^-$ can account for the observed
branching ratio of order $6\times 10^{-4}$. The branching ratio of
$B^-\to\Lambda_c\,\bar p\rho^-$ is larger than that of
$B^-\to\Lambda_c\,\bar p\pi^-$ by a factor of about 2.6\,. We
explain why the three-body charmful baryonic $B$ decay has a
larger rate than the two-body one, contrary to the case of mesonic
$B$ decays.

}

\pagebreak

\section{Introduction}
Inspired by the claim of the observation of the decay modes $p\bar
p\pi^\pm$ and $p\bar p\pi^+\pi^-$ in $B$ decays by ARGUS
\cite{ARGUS} in the late 1980s, baryonic $B$ decays were studied
extensively around the early 1990s
\cite{DTS,Paver,Gronau,Korner,Dobr,Chernyak,He,Lu,Jarfi,Ball,Khanna,Kaur}
with the focus on the two-body decay modes, e.g. $B\to p\bar
p,~\Lambda\bar\Lambda$. Up to now, none of the two-body baryonic
$B$ decays have been observed. Indeed, most of the earlier
predictions based on the pole model or QCD sum rule or the diquark
model are too large compared to experiment \cite{CLEOa,Belle} (see
Table I).

\begin{table}[ht]
\caption{Predictions of branching ratios for some two-body
baryonic $B$ decays in various models. We have normalized the
branching ratios to $|V_{ub}/V_{cb}|=0.085\,$. The predictions
given in [11] are carried out in two different quark-pair-creation
models: local and nonlocal. Experimental limits are taken from
[14,15].}
\begin{center}
\begin{tabular}{l| c c c c c c c }
&  &  &  &  & \multicolumn{2}{c}{[11]} &    \\ \cline{6-7}
\raisebox{1.5ex}[0cm][0cm]{} &
\raisebox{1.5ex}[0cm][0cm]{\cite{Paver}} &
\raisebox{1.5ex}[0cm][0cm]{\cite{Chernyak}} &
\raisebox{1.5ex}[0cm][0cm]{\cite{Lu}} &
\raisebox{1.5ex}[0cm][0cm]{\cite{Jarfi}} & non-local & local &
 \raisebox{1.5ex}[0cm][0cm]{experiment}  \\ \hline
 $\ov B^0\to\Lambda_c\bar p$ & & $4\times 10^{-4}$ &  $8.5\times
 10^{-4}$ & $1.1\times 10^{-3}$ & $1.7\times 10^{-3}$ & $1.9\times 10^{-3}$
& $<2.1\times  10^{-4}$ \\
 $\ov B^0\to p\bar p$ & $4.2\times 10^{-6}$ & $1.2\times 10^{-6}$ & $5.9\times
 10^{-6}$ & $7.0\times 10^{-6}$ & $2.9\times 10^{-6}$ & $2.7\times 10^{-5}$
& $<1.6\times  10^{-6}$ \\
 $\ov B^0\to\Lambda\bar\Lambda$ & & &  $3.0\times 10^{-6}$ & & $1.2\times 10^{-6}$ &
$2.3\times 10^{-5}$  &  $<2.3\times 10^{-6}$ \\
 $B^-\to\bar \Delta^{--} p$ & $1.5\times 10^{-4}$ & $2.9\times 10^{-7}$
  &  & $3.2\times 10^{-4}$ & $7.2\times 10^{-7}$ & $ 8.7\times 10^{-6}$
 &  $<1.5\times 10^{-4}$ \\
\end{tabular}
\end{center}
\end{table}

In order to understand why the momentum spectrum of produced
$\Lambda_c$ in inclusive $B$ decays is soft and why the two-body
decay modes, e.g. $\ov B\to\{\Lambda_c,\Sigma_c\}\{\bar
p,\bar\Delta\}$, have not been observed, Dunietz \cite{Dunietz}
argued that a straightforward Dalitz plot for the dominant $b\to
c\bar ud$ transition predicts the $c\,d$ invariant mass to be very
large. The very massive $c\,d\,q$ objects would be usually seen as
$\Lambda_c n\pi(n\geq 1)$ if the $c\,d$ forms a charmed baryon.
This explains the observed soft $\Lambda_c$ momentum spectrum and
the non-observation of $\Lambda_c\bar p$ decay. Since the very
massive $c\,d\,q$ could also be seen as $D^{(*)}NX$, the baryonic
processes $\ov B\to D^{(*)}N\bar N'X$ would be likely sizable.
Indeed, CLEO has recently reported the observation of $B^0\to
D^{*-}p\bar n$ at the $10^{-3}$ level and $B^0\to D^{*-}p\bar
p\pi^+$ at the $10^{-4}$ level \cite{CLEOb}. Theoretically, the
three-body decay modes $B\to D^{*-}N\bar N$ and $B^0\to
\rho^-(\pi^-)p\bar n$ have been recently studied in
\cite{CHT1,CHT2}.

A similar observation has been made by Hou and Soni \cite{HS}.
They pointed out that the smallness of the two-body baryonic decay
$B\to\B_1\ov\B_2$ has to do with the large energy release. They
conjectured that in order to have larger baryonic $B$ decays, one
has to reduce the energy release and  at the same time allow for
baryonic ingredients to be present in the final state. Under this
argument, the three-body decay, for example $B\to \rho p\bar n$,
will dominate over the two-body mode $B\to p\bar p$ since the
ejected $\rho$ meson in the former decay carries away much energy
and the configuration is more favorable for baryon production
because of reduced energy release compared to the latter
\cite{CHT2}. This is in contrast to the mesonic $B$ decays where
two-body decay rates are usually comparable to the three-body
modes. The large rate of $B^0\to D^{*-}p\bar n$ and $B^0\to
D^{*-}p\bar p\pi^+$ observed by CLEO indicates that the decays
$B\to$ baryons receive comparable contributions from $\ov
B\to\Lambda_c \bar p X$ and $\ov B\to [D]N\bar N' X$, where $[D]$
denotes any charmed meson. By the same token, it is expected that
for charmless baryonic $B$ decays, $\ov B\to \rho(\pi)\B_1\ov\B_2$
are the dominant modes induced by tree operators and $\ov
B\to(\pi,\eta',\rho)\B_{1(s)}\ov B_2$, e.g. $\ov B\to \rho\Lambda
\bar p$, are the leading modes induced by penguin diagrams.

In this work we focus on charmful baryonic decays $\ov
B\to\Lambda_c\bar p X$. The experimental results are summarized as
 \cite{PDG}:
 \be \label{data}
 \B(B^-\to\Lambda_c\bar p\pi^-)=(6.2\pm 2.7)\times 10^{-4}, &&
\qquad \B(\ov B^0\to\Lambda_c\bar p\pi^0)<5.9\times 10^{-4}, \non \\
 \B(\ov B^0\to\Lambda_c\bar p\pi^+\pi^-)=(1.3\pm 0.6)\times
10^{-3}, && \qquad \B(B^-\to\Lambda_c\bar p\pi^-\pi^0)<3.12\times
10^{-3},
 \en
together with the upper limit $\B(\ov B^0\to \Lambda_c\bar
p)<2.1\times 10^{-4}$. It is evident that the two-body mode is
suppressed. Specifically, we shall study $\ov B^0\to \Lambda_c
\,\bar p$ and $B^-\to \Lambda_c \,\bar p\,\pi^-(\rho^-)$ in detail
in order to understand their underlying decay mechanism. It has
been advocated that the $B$ decay to $\Lambda_c\bar p+\pi$'s is
suppressed relative to $\Lambda_c\bar p$ \cite{Jarfi}. We shall
see that this is not the case.

The layout of the present paper is organized as follows. In Sec.
II we first study the two-body charmful decay $\ov B^0\to
\Lambda_c \,\bar p$ to update the prediction of its branching
ratio. We then turn to the three-body decays $\ov B\to \Lambda_c
\,\bar p\,\pi(\rho)$ in Sec. III. A detail of the MIT bag model
for the evaluation of baryon-to-baryon weak transition matrix
elements is presented in the Appendix.

\vskip 0.3cm
\section{Two-body charmful baryonic decay $\ov B^0\to\Lambda_{\lowercase{c}}
\,\bar {\lowercase{p}}$} We first study the two-body baryonic
decay $\ov B^0\to\Lambda_c\,\bar p$  to update its prediction and
understand why it is suppressed compared to three-body modes. To
proceed, we first write down the relevant Hamiltonian
 \be \label{hamiltonian}
 {\cal H}_{\rm eff} &=& {G_F\over\sqrt{2}}
V_{cb}V_{ud}^*[c_1(\mu)O_1(\mu)+c_2(\mu)O_2(\mu)]+h.c.,
 \en
where $O_1=(\bar cb)(\bar d u)$ and $O_2=(\bar cu)(\bar db)$ with
$(\bar q_1q_2)\equiv \bar q_1\gamma_\mu(1-\gamma_5)q_2$. In order
to ensure that the physical amplitude is renormalization scale and
$\gamma_5$-scheme independent, we include vertex corrections to
hadronic matrix elements. This amounts to modifying the Wilson
coefficients by \cite{CCTY}:
 \be \label{effWC}
 c_1(\mu)\to c_1^{\rm eff} &=&
 c_1(\mu)+{\alpha_s\over 4\pi}\left(\gamma^{(0)T}\ln{m_b\over
 \mu}+\hat r^T\right)_{1i}c_i(\mu), \non \\
  c_2(\mu) \to c_2^{\rm eff} &=& c_2(\mu)+{\alpha_s\over 4\pi}
  \left(\gamma^{(0)T}\ln{m_b\over \mu}+\hat
r^T\right)_{2i}c_i(\mu),
 \en
where the anomalous dimension matrix $\gamma^{(0)}$ and the
constant matrix $\hat r$ in the naive dimensional regularization
and 't Hooft-Veltman schemes can be found in \cite{CCTY}. The
superscript $T$ in Eq. (\ref{effWC}) denotes a transpose of the
matrix. Numerically we have $c_1^{\rm eff}=1.168$ and $c_2^{\rm
eff}=-0.365$ \cite{CCTY}. It should be stressed that $c_1^{\rm
eff}$ and $c_2^{\rm eff}$ are renormalization scale and scheme
independent.\footnote{For the mesonic decay $B\to M_1M_2$ with two
mesons in the final state, two of the four quarks involving in the
vertex diagrams will form an ejected meson. In this case, it is
necessary to take into account the convolution with the ejected
meson wave function.} For later purposes we write
 \be \label{Ham}
 {\cal H}_{\rm eff} &=& {G_F\over\sqrt{2}}
 V_{cb}V_{ud}^*[c_+O_++c_-O_-]+h.c.,
 \en
with $O_\pm=O_1\pm O_2$ and $c_\pm={1\over 2}(c_1^{\rm eff}\pm
c_2^{\rm eff})$.

The decay amplitude of $\ov B^0\to\Lambda_c\,\bar p$ consists of
factorizable and nonfactorizable parts:
 \be
 A(\ov B^0\to\Lambda_c\,\bar p)=A(\ov B^0\to\Lambda_c\,\bar p)_{\rm fact}
 +A(\ov B^0\to\Lambda_c\,\bar p)_{\rm nonfact},
 \en
with
 \be
 A(\ov B^0\to\Lambda_c\,\bar p)_{\rm
 fact}={G_F\over\sqrt{2}}V_{cb}V_{ud}^*\,a_2\la \Lambda_c \bar p|(\bar
 cu)|0\ra\la 0|(\bar d b)|\ov B^0\ra,
 \en
where $a_2=c_2^{\rm eff}+c_1^{\rm eff}/N_c$. The short-distance
factorizable contribution is nothing but the $W$-exchange diagram.
This $W$-exchange contribution has been estimated and is found to
be very small and hence can be neglected \cite{Korner,Kaur}.
However, a direct evaluation of nonfactorizable contributions is
very difficult. This is the case in particular for baryons, which
being made out of three quarks, in contrast to two quarks for
mesons, bring along several essential complications. In order to
circumvent this difficulty, it is customary to assume that the
nonfactorizable effect is dominated by the pole diagram with
low-lying baryon intermediate states; that is, nonfactorizable
$s$- and $p$-wave amplitudes are dominated by ${1\over 2}^-$
low-lying baryon resonances and ${1\over 2}^+$ ground-state
intermediate states, respectively \cite{Jarfi}. For $\ov
B^0\to\Lambda_c\,\bar p$, we consider the strong-interaction
process $\ov B^0\to \Sigma_b^{+(*)}\,\bar p$ followed by the weak
transition $\Sigma_b^{+(*)}\to \Lambda_c$, where $\Sigma_b^*$ is a
${1\over 2}^-$ baryon resonance (see Fig. 1). Considering the
strong coupling
 \be
 ig_{B p\Sigma_b}\ov\psi_{\Sigma_b}\gamma_5\psi_{p}\phi_B+g_{B
 p\Sigma_b^*}\ov\psi_{\Sigma_b^*}\psi_p\phi_B,
 \en
the pole-diagram amplitude has the form
 \be
 A(\ov B^0\to\Lambda_c\bar p)_{\rm nonfact}=\bar u_{\Lambda_c}(A+B\gamma_5)v_p,
 \en
where
 \be
 A=-{g_{B^0p\Sigma_b^{+*}}b_{\Sigma_b^*\Lambda_c}\over
 m_{\Lambda_c}-m_{\Sigma_b^*} }, \qquad\qquad B=\,{g_{B^0p\Sigma_b^+}
 a_{\Sigma_b\Lambda_c}\over
 m_{\Lambda_c}-m_{\Sigma_b}}
 \en
correspond to $s$-wave parity-violating (PV) and $p$-wave
parity-conserving (PC) amplitudes, respectively, and
 \be
 \la \Lambda_c|{\cal H}_{\rm eff}^{\rm PC}|\Sigma_b^+\ra = \bar
 u_{\Lambda_c}a_{\Sigma_b\Lambda_c}u_{\Sigma_b}, \qquad\quad
 \la \Lambda_c|{\cal H}_{\rm eff}^{\rm PV}|\Sigma_b^{*+}\ra = i\bar
 u_{\Lambda_c}b_{\Sigma_b^*\Lambda_c}u_{\Sigma_b^*}.
 \en

\begin{figure}[tb]
\hspace{2cm}\psfig{figure=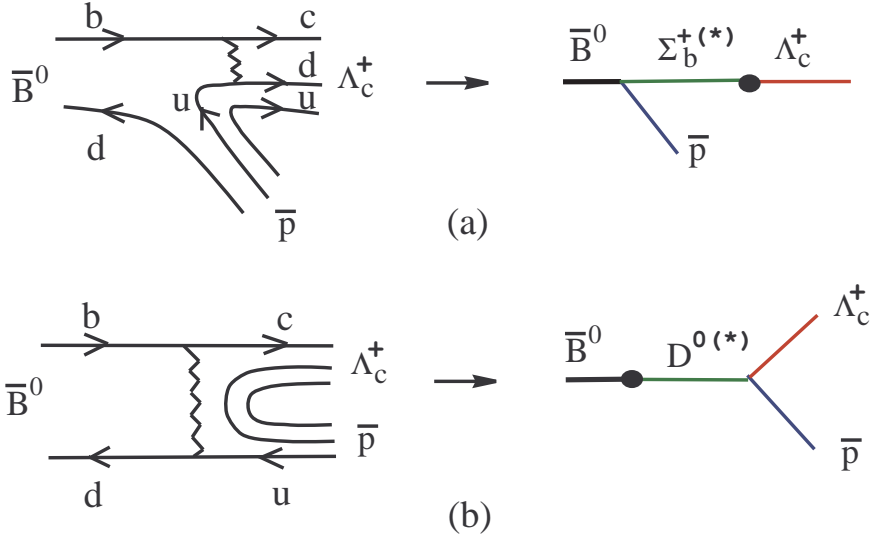,height=3in} \vspace{1.2cm}
    \caption{{\small Some pole diagrams for $\ov B^0\to\Lambda_c\bar
    p$ where the symbol $\bullet$ denotes the weak vertex. (a) corresponds
    to nonfactorizable internal
    $W$ emission, while (b) to the $W$-exchange contribution.
    }}
   \label{fig:1}
\end{figure}

The main task is to evaluate the weak matrix elements and the
strong coupling constants. We shall employ the MIT bag model
\cite{MIT} to evaluate the baryon matrix elements (see e.g.
\cite{CT92,CT93} for the method). Since the quark-model wave
functions best resemble the hadronic states in the frame where
both baryons are static, we thus adopt the static bag
approximation for the calculation. Note that because the
four-quark operator $O_+$ is symmetric in color indices, it does
not contribute to the baryon-baryon matrix element since the
baryon-color wave function is totally antisymmetric. From Eq.
(\ref{Ham}) and the Appendix we obtain the PC matrix element
 \be \label{PCm.e.}
 a_{\Sigma_b\Lambda_c}=-{G_F\over
 \sqrt{2}}V_{cb}V_{ud}^*\,c_-{4\over\sqrt{6}}(X_1+3X_2)(4\pi),
 \en
where
 \be \label{bagX}
 X_1 &=& \int^R_0
 r^2dr[u_c(r)v_u(r)-v_c(r)u_u(r)][u_d(r)v_b(r)-v_d(r)u_b(r)], \non
 \\
 X_2 &=& \int^R_0
 r^2dr[u_c(r)u_u(r)+v_c(r)v_u(r)][u_d(r)u_b(r)+v_d(r)v_b(r)],
 \en
are four-quark overlap bag integrals (see the Appendix for
notation). In principle, one can also follow \cite{CT92} to tackle
the low-lying negative-parity $\Sigma_b^*$ state in the bag model
and evaluate the PV matrix element
$b_{\Sigma^*_c\Lambda_c}$.\footnote{In the bag model the low-lying
negative parity baryon states are made of two quarks in the ground
$1S_{1/2}$ eigenstate and one quark excited to $1P_{1/2}$ or
$1P_{3/2}$. Consequently, the evaluation of the PC matrix element
for ${1\over 2}^--{1\over 2}^+$ baryonic transition  becomes much
involved owing to the presence of $1P_{1/2}$ and $1P_{3/2}$ bag
states.} However, it is known that the bag model is less
successful even for the physical non-charm and non-bottom ${1\over
2}^-$ resonances \cite{MIT}, not mentioning the charm or bottom
${1\over 2}^-$ resonances. In short, we know very little about the
${1\over 2}^-$ state. Therefore, we will not evaluate the PV
matrix element $b_{\Sigma_b^*\Lambda_c}$ as its calculation in the
bag model is much involved  and is far more uncertain than the PC
one \cite{CT92}.

Using the bag wave functions given in the Appendix,  we find,
numerically,
 \be
 X_1=-1.49\times 10^{-5}\,{\rm GeV}^3, \qquad\quad
 X_2=1.81\times 10^{-4}\,{\rm GeV}^3.
 \en
The decay rate of $B\to\B_1\ov \B_2$ is given by
 \be
 \Gamma(B\to \B_1\ov \B_2)&=& {p_c\over 4\pi}\Bigg\{
 |A|^2\,{(m_B+m_1+m_2)^2p_c^2\over (E_1+m_1)(E_2+m_2)m_B^2}\non \\
 & +&
 |B|^2\,{[(E_1+m_1)(E_2+m_2)+p_c^2]^2\over
 (E_1+m_1)(E_2+m_2)m_B^2} \Bigg\},
 \en
where $p_c$ is the c.m. momentum, and $E_i$ and $m_i$ are the
energy and mass of the baryon $\B_i$, respectively. Putting
everything together we obtain
 \be
 \B(\ov B^0\to\Lambda_c\bar p)_{\rm PC}=7.2\times
 10^{-6}\left|{g_{B^0p\Sigma_b^+}\over 6}\right|^2.
 \en
The PV contribution is expected to be smaller. For example, it is
found to be $\Gamma^{\rm PV}/\Gamma^{\rm PC}=0.59$ in
\cite{Jarfi}. Therefore, we conclude that
 \be \label{Lamcp}
 \B(\ov B^0\to\Lambda_c\bar p)\lsim 1.1\times
 10^{-5}\left|{g_{B^0p\Sigma_b^+}\over 6}\right|^2.
 \en
The strong coupling $g_{B^0p\Sigma_b^+}$ has been estimated in
\cite{Jarfi} using the quark-pair-creation model and it is found
to lie in the range $g_{B^0p\Sigma_b^+}=-(6\sim 10)$, recalling
that $g_{\pi NN}\approx 14$. We shall see in Sec. III.A that the
measurement of $B^-\to\Lambda_c\bar p\pi^-$ can be used to extract
the coupling $g_{B^+p\Lambda_b}$ which in turn provides
information on $g_{B^0p\Sigma_b^+}$. At any rate, the prediction
(\ref{Lamcp}) is consistent with the current experimental limit
$2.1\times 10^{-4}$ \cite{PDG}. Note that all earlier predictions
based on the QCD sum rule \cite{Chernyak} or the pole model
\cite{Jarfi} or the diquark model \cite{Ball} are too large
compared to experiment (see Table I). In the pole-model
calculation in \cite{Jarfi}, the weak matrix element is largely
overestimated.

\vskip 0.3cm
\section{Three-body charmful baryonic decays}
\subsection{$B^-\to \Lambda_{\lowercase{c}}
\,\bar {\lowercase{p}}\,\pi^-$}
 The quark diagrams and the corresponding pole diagrams for $B^-\to \Lambda_c
\,\bar p\,\pi^-$ are shown in Fig. 2. There exist two distinct
internal $W$ emissions and only one of them is factorizable,
namely Fig. 2(b). The external $W$ emission diagram Fig. 2(a) is
of course factorizable. Therefore, unlike the two-body decay $\ov
B^0\to\Lambda_c\bar p$, the three-body mode $B^-\to \Lambda_c
\,\bar p\,\pi^-$ does receive sizable factorizable contributions
 \be \label{factamp}
 A(B^-\to\Lambda_c\bar p\pi^-)_{\rm fact} &=&
 {G_F\over\sqrt{2}}V_{cb}V_{ud}^*\Big\{a_1\la \pi^-|(\bar
 du)|0\ra\la \Lambda_c\bar p|(\bar cb)|B^-\ra \non\\
 &+& a_2\la\pi^-|(\bar db)|B^-\ra\la \Lambda_c\bar p|(\bar
 cu)|0\ra\Big\} \equiv  A_1+A_2,
 \en
where naively $a_1=c_1^{\rm eff}+c_2^{\rm eff}/N_c$ and
$a_2=c_2^{\rm eff}+c_1^{\rm eff}/N_c$, to which we will come back
later.  Unfortunately, in practice we do not know how to evaluate
the 3-body hadronic matrix element $\la\Lambda_c\bar p|(\bar
cb)|B^-\ra$. Thus we will instead evaluate the corresponding
low-lying pole diagrams for external $W$-emission, namely, the
strong process $B^-\to \Lambda_b^{(*)}\bar p$, followed by the
weak decay $\Lambda_b^{(*)}\to\Lambda_c\pi^-$ [see Fig. 2(a)]. Its
amplitude is given by
 \be
 A_1 &=& -{G_F\over\sqrt{2}}V_{ud}V_{cb}^*\,g_{B^+p\Lambda_b}f_\pi\,a_1\,\bar
 u_{\Lambda_c}\Big\{f_1^{\Lambda_b\Lambda_c}(m_\pi^2)[2p_\pi\cdot
 p_{\Lambda_c}+p\!\!\!/_\pi(m_{\Lambda_b}-m_{\Lambda_c})]\gamma_5  \non \\
 && +g_1^{\Lambda_b\Lambda_c}(m_\pi^2)[2p_\pi\cdot
 p_{\Lambda_c}-p\!\!\!/_\pi(m_{\Lambda_b}+m_{\Lambda_c})]\Big\}
 v_{\bar p}\times{1\over
 (p_{\Lambda_c}+p_\pi)^2-m_{\Lambda_b}^2 },
 \en
where we have applied factorization to the weak decay
$\Lambda_b\to\Lambda_c\pi$ and employed the form factors defined
by
 \be
 \la\Lambda_c^+(p_{\Lambda_c})|(\bar cb)|\Lambda_b^0(p_{\Lambda_b})\ra &=&
\bar
u_{\Lambda_c}\Bigg\{f_1^{\Lambda_b\Lambda_c}(p_\pi^2)\gamma_\mu
+i{f_2^{\Lambda_b\Lambda_c}(p_\pi^2)\over m_{\Lambda_b} }
 \sigma_{\mu\nu}p_\pi^\nu+{f_3^{\Lambda_b\Lambda_c}(p_\pi^2)\over m_{\Lambda_b}}p_{\pi\mu} \non
 \\ &&
 -[g_1^{\Lambda_b\Lambda_c}(p_\pi^2)\gamma_\mu+i{g_2^{\Lambda_b\Lambda_c}(p_\pi^2)\over
 m_{\Lambda_b}}
 \sigma_{\mu\nu}p_\pi^\nu+{g_3^{\Lambda_b\Lambda_c}(p_\pi^2)\over m_{\Lambda_b}}p_{\pi\mu}]
\gamma_5\Bigg\}u_{\Lambda_b},
 \en
where $p_\pi=p_{\Lambda_b}-p_{\Lambda_c}$. Note that the ${1\over
2}^-$ intermediate state $\Lambda_b^*$ makes no contribution as
the matrix element $\la \Lambda_c|(\bar cb)|\Lambda_b^*\ra$
vanishes. Likewise, the intermediate states $\Sigma_b^0$ and
$\Sigma_b^{0*}$ also do not contribute to $A_1$ under the
factorization approximation because the weak transition  $\la
\Lambda_c|(\bar cb)|\Sigma_b^{0(*)}\ra$ is prohibited as
$\Sigma_b$ and $\Sigma_b^{*}$ are sextet bottom baryons whereas
$\Lambda_c$ is a anti-triplet charmed baryon.

\begin{figure}[p]
\hspace{1cm}\centerline{\psfig{figure=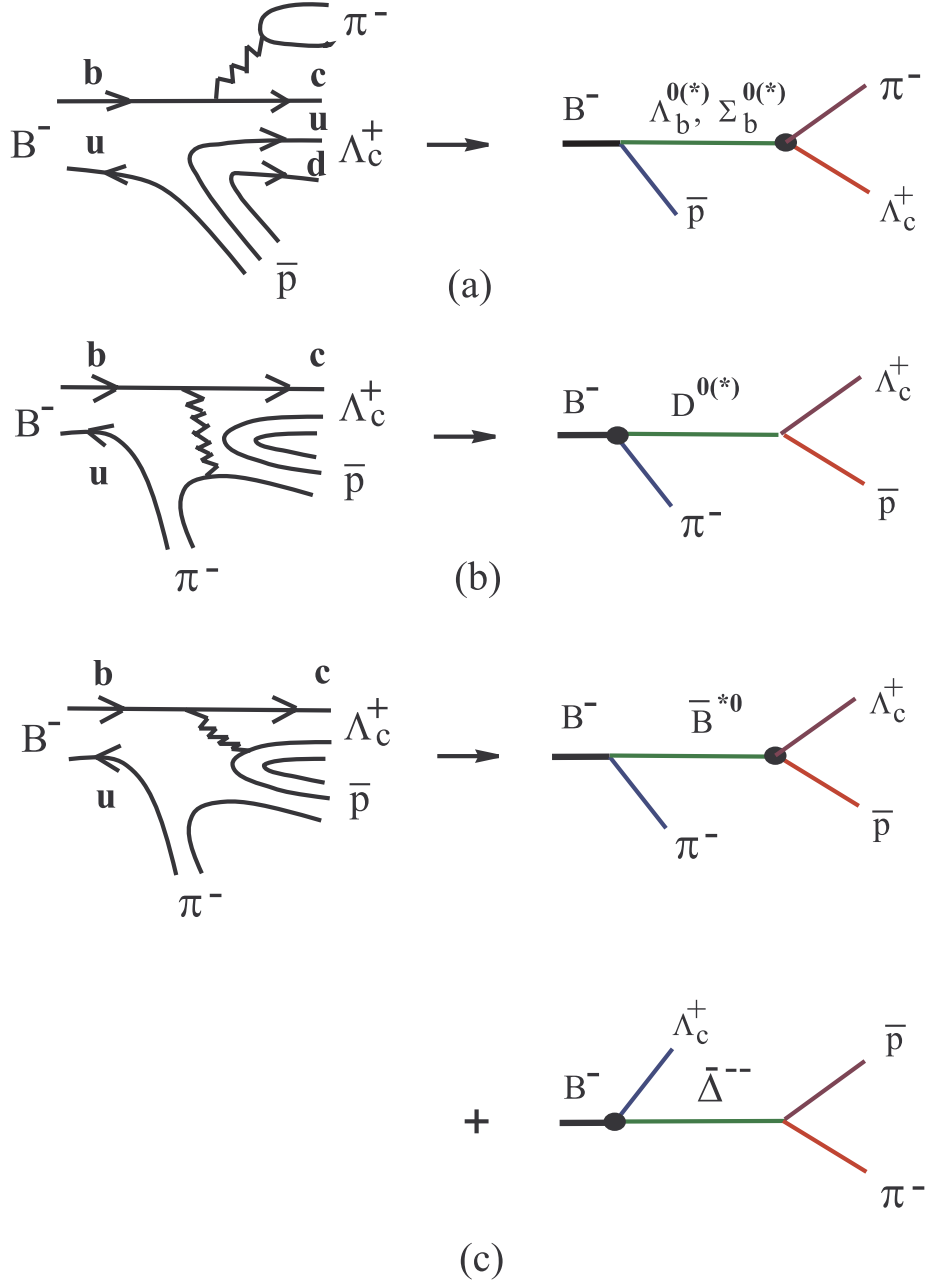,height=7in}}
\vspace{0.5cm}
    \caption{{\small Some pole diagrams for $B^-\to\Lambda_c\bar p\pi^-$
    where the symbol $\bullet$ denotes the weak vertex.
    (a) and (b) correspond to factorizable external and internal $W$-emission
    contributions, respectively, while (c) to
    nonfactorizable internal $W$-emission diagrams. There are two pole diagrams
    corresponding to the quark diagram in (c).
    }}
   \label{fig:2}
\end{figure}

To evaluate the fcatorizable amplitude $A_2$, as shown in
Fig.~2(b), we apply the parametrization for the $B-\pi$ matrix
element
 \be
 \la\pi^-(p_\pi)|(\bar
 db)|B^-(p_B)\ra=F_1^{B\pi}(q^2)(p_B+p_\pi)_\mu+\left(F_0^{B\pi}
 (q^2)-F_1^{B\pi}(q^2)\right){m_B^2-m_\pi^2\over q^2}q_\mu,
 \en
and obtain
 \be
 A_2={G_F\over\sqrt{2}}V_{ud}V_{cb}^*\,a_2\bar
 u_{\Lambda_c}\left[(ap\!\!\!/_\pi+b)-(cp\!\!\!/_\pi+d)\gamma_5\right]v_p,
 \en
where
 \be
 a&=&2f_1^{\Lambda_cp}(t)F_1^{B\pi}(t)+2f_2^{\Lambda_cp}(t)F_1^{B\pi}(t)
(m_{\Lambda_c}+m_p)/m_{\Lambda_c}, \non \\
 b&=& (m_{\Lambda_c}-m_p)f_1^{\Lambda_cp}(t)\left[F_1^{B\pi}(t)+(F^{B\pi}_0(t)
 -F^{B\pi}_1(t)){m_B^2-m_\pi^2\over t}\right]  \non \\
 &&-2f_2^{\Lambda_cp}(t)F_1^{B\pi}(t)(p_{\Lambda_c}-p_p)\cdot p_\pi/m_{\Lambda_c}
 +f_3^{\Lambda_cp}(t)F_0^{B\pi}(t)(m_B^2-m_\pi^2)/m_{\Lambda_c}, \non \\
 c &=& 2g_1^{\Lambda_cp}(t)F_1^{B\pi}(t)+2g_2^{\Lambda_cp}(t)F_1^{B\pi}(t)
(m_{\Lambda_c}-m_p)/m_{\Lambda_c}, \non \\
 d&=&(m_{\Lambda_c}+m_p)g_1^{\Lambda_cp}(t)\left[F_1^{B\pi}(t)+
 (F^{B\pi}_0(t)-F^{B\pi}_1(t)){m_B^2-m_\pi^2\over
 t}\right]  \non \\
 &&-2g_2^{\Lambda_cp}(t)F_1^{B\pi}(t)(p_{\Lambda_c}-p_p)\cdot
 p_\pi/m_{\Lambda_c}+g_3^{\Lambda_cp}(t)F_0^{B\pi}(t)(m_B^2-m_\pi^2)/m_{\Lambda_c},
 \en
and $t\equiv q^2=(p_B-p_\pi)^2=(p_{\Lambda_c}+p_p)^2$.

The form factors $f_i$ and $g_i$ for the heavy-to-heavy and
heavy-to-light baryonic transitions at zero recoil have been
computed using the non-relativistic quark model \cite{CT96}. In
principle, HQET puts some constraints on these form factors.
However, it is clear that HQET is not adequate for our purposes:
the predictive power of HQET for the baryon form factors at order
$1/m_Q$ is limited only to the antitriplet-to-antitriplet heavy
baryonic transition. Hence, we will follow \cite{CT96} to apply
the nonrelativistic quark model to evaluate the weak
current-induced baryon form factors at zero recoil in the rest
frame of the heavy parent baryon, where the quark model is most
trustworthy. This quark model approach has the merit that it is
applicable to heavy-to-heavy and heavy-to-light baryonic
transitions at maximum $q^2$ and that it becomes meaningful to
consider $1/m_q$ corrections as long as the recoil momentum is
smaller than the $m_q$ scale. It has been shown in \cite{CT96}
that the quark model predictions agree with HQET for the
antitriplet-to-antitriplet (e.g., $\Lambda_b\to\Lambda_c,~
\Xi_b\to\Xi_c$) form factors to order $1/m_Q$. For sextet
$\Sigma_b\to \Sigma_c$ and $\Omega_b\to\Omega_c$ transitions, the
quark-model results  are also in accord with the HQET predictions
(for details see \cite{Cheng97}). Numerically we have
\cite{Cheng97}
 \be
&&
f_1^{\Lambda_b\Lambda_c}(q^2_m)=g_1^{\Lambda_b\Lambda_c}(q^2_m)=1.02,
 \quad
 f_2^{\Lambda_b\Lambda_c}(q^2_m)=g_3^{\Lambda_b\Lambda_c}(q^2_m)=-0.23,
 \non \\
&&
f_3^{\Lambda_b\Lambda_c}(q^2_m)=g_2^{\Lambda_b\Lambda_c}(q^2_m)=-0.03,
 \en
for the $\Lambda_b-\Lambda_c$ transition at zero recoil
$q_m^2=(m_{\Lambda_b}-m_{\Lambda_c})^2$, and \cite{CT96}
 \be
&& f_1^{\Lambda_cp}(q^2_m)=g_1^{\Lambda_cp}(q^2_m)=0.80,
 \quad
 f_2^{\Lambda_cp}(q^2_m)=g_3^{\Lambda_cp}(q^2_m)=-0.21,
 \non \\
 && f_3^{\Lambda_cp}(q^2_m)=g_2^{\Lambda_cp}(q^2_m)=-0.07,
 \en
for the $\Lambda_c-p$ transition at $q^2_m=(m_{\Lambda_c}-m_p)^2$.

Since the calculation for the $q^2$ dependence of form factors is
beyond the scope of the non-relativistic quark model, we will
follow the conventional practice to assume a pole dominance for
the form-factor $q^2$ behavior:
 \be
 f(q^2)=f(q^2_m)\left({1-q^2_m/m^2_V\over 1-q^2/m_V^2} \right)^n\,,\qquad
 g(q^2)=g(q^2_m)
\left({1-q^2_m/m^2_A\over 1-q^2/m_A^2} \right)^n\,,
 \en
where $m_V$ ($m_A$) is the pole mass of the vector (axial-vector)
meson with the same quantum number as the current under
consideration. The function
 \be
 G(q^2)=\left({1-q^2_m/m^2_{\rm
pole}\over 1-q^2/m_{\rm pole}^2} \right)^n
 \en
plays the role of the baryon Isgur-Wise function $\zeta(\omega)$
for the $\Lambda_Q\to \Lambda_{Q'}$ transition, namely, $G=1$ at
$q^2=q^2_m$. The function $\zeta(\omega)$ has been calculated in
the literature in various different models
\cite{Jenkins,Sadzi,GuoK,Dai,Iva97,Iva99}. Using the pole masses
$m_V=6.34$ GeV, $m_A=6.73$ GeV for the $\Lambda_b\to\Lambda_c$
transition, it is found that $G(q^2)$ is consistent with the
earlier soliton model \cite{Jenkins} and MIT bag model
\cite{Sadzi} calculation of $\zeta(\omega)$ for $n=2$ \cite{CT96}.
However, a recent calculation of $\zeta(\omega)$ in \cite{Iva99}
yields
 \be
 \zeta(\omega)=\left({2\over 1+\omega}\right)^{1.23+0.4/\omega}
 \en
and this favors $n=1$. Therefore, whether the $q^2$ dependence is
monopole or dipole for heavy-to-heavy transitions is not clear.
Hence we shall use both monopole and dipole dependence in ensuing
calculations. Moreover, one should bear in mind that the $q^2$
behavior of form factors is probably more complicated and it is
likely that a simple pole dominance only applies to a certain
$q^2$ region, especially for the heavy-to-light transition. For
the $\Lambda_c-p$ transition, we will use the pole masses
$m_V=2.01$ GeV and $m_A=2.42$ GeV and assume dipole $q^2$
dependence.

For the form factors $F_{0,1}^{B\pi}(q^2)$ we consider the
recently proposed Melikhov-Stech (MS) model based on the
constituent quark picture \cite{MS}. Although the form factor
$q^2$ dependence is in general model dependent, it should be
stressed that $F_1^{B\pi}(q^2)$ increases with $q^2$ more rapidly
than $F_0^{B\pi}(q^2)$ as required by heavy quark symmetry. We
shall see below that the predicted decay rates are insensitive to
the choice of form-factor models.

Thus far we have only discussed factorizable contributions. The
nonfactorizable effects are conventionally estimated by evaluating
the corresponding pole diagrams. The processes
 \be
 B^-\to \pi^- + &\ov B^{*0}& \non \\
 &\hookrightarrow& \bar p+\Lambda_c \non \\
 B^-\to \Lambda_c+ &\bar\Delta^{--}& \non \\
 &\hookrightarrow& \bar p+\pi^-
 \en
are some examples of the pole diagrams shown in Figs. 2(c) and
2(d); they correspond to nonfactorizable internal $W$-emission.
Presumably these nonfactorizable contributions will affect the
parameter $a_2$ substantially.

The total decay rate for the process $B^-(p_B)\to
\Lambda_c(p_1)+\bar p(p_2)+\pi^-(p_3)$ is computed by
 \be \Gamma = {1\over (2\pi)^3}\,{1\over
32m_B^3}\int |A|^2dm_{12}^2dm_{23}^2,
 \en or
  \be
\Gamma={1\over (2\pi)^3}\,{1\over 16m_B^2}\int |A|^2dE_\pi
dm_{23}^2,
 \en
where $E_\pi$ is the energy of the outgoing pion, and
$m_{ij}^2=(p_i+p_j)^2$ with $p_3=p_\pi$. For a given $E_\pi$, the
range of $m_{23}^2$ is fixed by kinematics. Under naive
factorization, the parameter $a_2$ appearing in Eq.
(\ref{factamp}) is numerically equal to 0.024, which is very small
compared to the value of $a_2=0.40-0.55$ extracted from $\ov B^0
\to D^{0(*)}\pi^0$ decays \cite{Cheng01} and $|a_2|=0.26\pm 0.02$
in $B\to\J K$ decay \cite{a1a2}. As stated before, $a_2$ may
receive sizable contributions from nonfactorizable pole diagrams
Figs. 2(c) and 2(d). Therefore, we will treat $a_2$ as a free
parameter and take $a_2=0.30$ as an illustration. For strong
coupling constants a simple quark-pair-creation model yields (see
Appendix C of \cite{Jarfi} for detail)
 \be \label{coupling}
 |g_{B^+p\Lambda_b}|=3\sqrt{3/2}\,|g_{B^0p\Sigma_b^+}|.
 \en
Hence, the strong coupling constant $|g_{B^+p\Lambda_b}|$ is much
larger than $|g_{B^0 p\Sigma_b^+}|$. Putting everything together
we obtain numerically
 \be \label{BRpion}
 \B(B^-\to\Lambda_c\bar p\pi^-)&=&\cases{ (10.0r^2+0.04-0.8r)\times 10^{-4} & for~$n=1$ \cr
 (5.5r^2+0.04-0.6r)\times 10^{-4} & for~$n=2$ }  \non \\
 &=& \cases{ 9.2\times 10^{-4} & for~$n=1$~and~$g_{B^+p\Lambda_b}=16$ \cr
 4.9\times 10^{-4} & for~$n=2$~and~$g_{B^+p\Lambda_b}=16$ },
 \en
where $r=g_{B^+p\Lambda_b}/16$ and the first two lines show
explicitly the contributions from external $W$-emission, internal
$W$-emission and their interference, respectively.  We find that
the external $W$-emission and internal $W$-emission contribute
destructively (constructively) if the $\Lambda_c-p$ baryonic form
factor $q^2$ dependence is of the dipole (monopole) form. From Eq.
(\ref{BRpion}) we find that the strong coupling constant
$g_{B^+p\Lambda_b}$ in the vicinity of order 16 can accommodate
the observed branching ratio of $B^-\to\Lambda_c\bar p\pi^-$ [see
Eq. (\ref{data})]. It follows from Eq. (\ref{coupling}) that
$|g_{B^0p\Sigma_b^+}|\sim 4.3$, which is close to the model
estimate of $6\sim 10$ given in \cite{Jarfi}. It is likely that
the quark-pair-creation-model calculation of strong couplings is
more reliable for their ratios than their absolute values.

We have checked explicitly that the results are fairly insensitive
to the choice of $B-\pi$ form factors. For example, we have
computed the branching ratios using the three different
form-factor models given in \cite{formfactor} and found that the
difference in rates is at most at the level of 5\%.

Evidently, the calculated branching ratios are in agreement with
experiment (\ref{data}). There are several reasons why the
three-body decay rate of $B^-\to \Lambda_c\bar p\pi^-$ is larger
than that of the two-body one $\ov B^0\to \Lambda_c\bar p$. (i)
The former decay receives external and internal $W$-emission
contributions, whereas the color-suppressed factorizable
$W$-exchange contribution to the latter is greatly suppressed.
(ii) At the pole-diagram level, the $\Sigma_b$ propagator in the
pole amplitude for the latter is of order $1/(m_b^2-m_c^2)$, while
the invariant mass of the $(\Lambda_c\bar p)$ system can be large
enough in the former decay so that its propagator in the pole
diagram is not subject to the same $1/m_b^2$ suppression. (iii)
The strong coupling constant for $\Lambda_b\to B^- p$ is larger
than that for $\Sigma_b^+\to\ov B^0 p$.

\subsection{$B^-\to \Lambda_{\lowercase{c}} \,\bar
{\lowercase{p}}\,\rho^-$} Naively it is expected that
$\Lambda_c\bar p\rho^-$ has a larger rate than $\Lambda_c\bar
p\pi^-$ due to the three polarization states for the $\rho$ meson.
The calculation for $B^-\to\Lambda_c\,\bar p\,\rho^-$ is the same
as that for $B^-\to\Lambda_c\,\bar p\,\pi^-$ except that two of
the matrix elements are replaced by
 \be
 \la\rho^-|(\bar du)|0\ra &=& f_\rho m_\rho\vp_\mu^*,
 \en
and
 \be
 \la\rho^-|(\bar db)|B^-\ra &=& {2\over m_B+m_\rho}\epsilon_{\mu\nu\alpha\beta}
\vp^{*\nu}p^\alpha_B p^\beta_\rho V^{B\rho}(q^2)-i\Bigg\{ (m_B+m_\rho)
\vp^*_\mu A_1^{B\rho}(q^2) \non \\
&-& {\vp^*\cdot p_B\over m_B+m_\rho}(p_B+p_\rho)_\mu
A_2^{B\rho}(q^2)-2m_\rho {\vp^*\cdot p_B\over
q^2}q_\mu\left[A_3^{B\rho}(q^2)-A_0^{B\rho}(q^2)\right]\Bigg\},
 \en
where $q=p_B-p_\rho=p_{\Lambda_c}+p_p$ and
 \be
A_3^{B\rho}(q^2)={m_B+m_\rho\over
2m_\rho}A_1^{B\rho}(q^2)-{m_B-m_\rho\over
2m_\rho}A_2^{B\rho}(q^2).
 \en
Obviously the calculation is much more involved owing to the
presence of the four form factors $V,~A_0,~A_1,~A_2$ compared to
the pion case where there are only two form factors $F_0$ and
$F_1$.

A straightforward but tedious calculation yields
 \be
 \B(B^-\to\Lambda_c\,\bar p\,\rho^-)&=&
\cases{ (2.6r^2+0.02-0.3r)\times 10^{-3} & for~$n=1$ \cr
 (1.5r^2+0.02-0.2r)\times 10^{-3} & for~$n=2$}  \non \\
 &=&  \cases{ 2.3\times 10^{-3} & for~$n=1$~and~$g_{B^+p\Lambda_b}=16$ \cr
 1.3\times 10^{-3} & for~$n=2$~and~$g_{B^+p\Lambda_b}=16$ },
 \en
where we have used the decay constant $f_\rho=216$ MeV and the MS
model \cite{MS} for the $B-\rho$ form factors. As in the previous
case, the contributions from external $W$-emission, internal
$W$-emission and their interference are shown explicitly in the
first two lines of the above equation. Again we have checked
explicitly that the predictions are insensitive to the form-factor
models for the $B-\rho$ transition. Note that the predicted
branching ratio is consistent with the current limit on
$B^-\to\Lambda_c\bar p\pi^-\pi^0$ [see Eq.~(\ref{data})]. The
ratio
 \be
 {\Gamma(B^-\to\Lambda_c\,\bar p\,\rho^-)\over \Gamma(B^-\to\Lambda_c\,\bar
 p\,\pi^-)}= 2.6
 \en
for $n=1$ or 2 is independent of the strong coupling
$g_{B^+p\Lambda_c}$ and hence its prediction should be more
trustworthy. Experimentally it is important to search for the $B$
decay into $\Lambda_c\bar p\rho^-$ and have a refined measurement
of $\Lambda_c\bar p\pi^-$ in order to understand their underlying
decay mechanism.

\section{Conclusions}
We have studied the two-body and three-body charmful baryonic $B$
decays: $\ov B^0\to \Lambda_c \,\bar p$ and  $B^-\to \Lambda_c
\,\bar p\,\pi^-(\rho^-)$. The factorizable $W$-exchange
contribution to $\ov B^0\to\Lambda_c\,\bar p$ is negligible.
Applying the bag model to evaluate the weak $\Sigma_b-\Lambda_c$
transition, we find $\B(\ov B^0\to\Lambda_c\,\bar p)\lsim
1.1\times 10^{-5}|g_{B^0p\Sigma_b^+}/ 6|^2$ with
$g_{B^0p\Sigma_b^+}$ being a strong coupling for the decay
$\Sigma_b^+\to \ov B^0 p$ and the predicted branching ratio is
well below the current experimental limit $2.1\times 10^{-4}$.
Contrary to the two-body mode, the three-body decay
$B^-\to\Lambda_c\,\bar p\pi^-$ receives factorizable external and
internal $W$-emission contributions. The external $W$-emission
amplitude involves a three-body hadronic matrix element that
cannot be evaluated directly. Instead we consider the
corresponding pole diagram that mimics the external $W$-emission
at the quark level. It is found that the factorizable
contributions to $B^-\to\Lambda_c\,\bar p\pi^-$ can account for
the observed branching ratio of order $6\times 10^{-4}$. The
strong coupling $|g_{B^+p\Lambda_b}|$ is extracted to be of order
16, which in turn implies $|g_{B^0p\Sigma_b^+}|\sim 4.3$ under the
quark-pair-creation model assumption. The decay rate of
$B^-\to\Lambda_c\,\bar p\rho^-$ is larger than that of
$B^-\to\Lambda_c\,\bar p\pi^-$ by a factor of $2.6$. We have shown
and explained  why the 3-body charmful baryonic $B$ decay in
general has a larger rate than the 2-body one.

Finally, our present study is ready to generalize to other
charmful baryonic $B$ decays, e.g.
$B\to\Lambda_c\bar\Delta,~\Sigma_c\bar N$,
$B\to\Lambda_c\bar\Delta \pi(\rho),~\Sigma_c\bar N\pi(\rho),
\cdots,$ etc. Experimentally it would be interesting and important
to measure these hadronic decays.

\vskip 3.0cm \acknowledgments  One of us (H.Y.C.)  wishes to thank
Physics Department, Brookhaven National Laboratory and C.N. Yang
Institute for Theoretical Physics at SUNY Stony Brook for their
hospitality. This work was supported in part by the National
Science Council of R.O.C. under Grant Nos. NSC90-2112-M-001-047
and NSC90-2112-M-033-004.

\newpage \centerline{\bf APPENDIX}
\renewcommand{\thesection}{\Alph{section}}
\renewcommand{\theequation}{\thesection\arabic{equation}}
\setcounter{equation}{0} \setcounter{section}{1} \vskip 0.5 cm
\vskip 0.3cm

In this Appendix we evaluate the baryon matrix elements in the MIT
bag model \cite{MIT}.  In this model the quark spatial wave
function is given by
 \be
\psi_{_{S_{1/2}}} &=& \,{N_{-1}\over(4\pi
R^3)^{1/2}}\left(\matrix{ ij_0(xr/R)\chi \cr
-\sqrt{\epsilon}j_1(xr/R)\vec{\sigma}\cdot\hat{r}\chi\cr} \right)
\non \\  &\equiv & \left(\matrix{iu(r)\chi  \cr
v(r)\vec{\sigma}\cdot \hat{r}\chi \cr}\right)
 \en
for the quark in the ground $(1S_{1/2})$ state, where $j_0$ and
$j_1$ are spherical Bessel functions. The normalization factor
reads
 \be \label{normalization}
 N_{-1}= \,{x^2\over [2\omega(\omega-1)+mR]^{1/2}\sin x},
 \en
where $\epsilon=(\omega-mR)/(\omega+mR),~x=(\omega^2-m^2R^2)
^{1/2}$ for a quark of mass $m$ existing within a bag of radius
$R$ in mode $\omega$. For convenience, we have dropped in Eq.
(\ref{normalization}) the subscript $-1$ for $x,~\omega$ and $R$.
The eigenvalue $x$ is determined by the transcendental equation
 \be
 \tan x= \,{x\over 1-mR-(x^2+m^2R^2)^{1/2}}.
 \en

In terms of the large and small components $u(r)$ and $v(r)$ of
the $1S_{1/2}$ quark wave function, the matrix elements of the
two-quark operators $V_\mu(x)=\bar{q}'\gamma_\mu q$ and
$A_\mu(x)=\bar{q}'\gamma_ \mu\gamma_5q$ are given by
  \be \label{m.e.}
\la{q'}|V_0|{q}\ra &=&  u'u+v'v, \non \\    \la{q'}|A_0|{q}\ra
&=&-i (u'v-v'u)\vec{\sigma}\cdot\hat{r}, \non  \\
\la{q'}|\vec{V}|{q}\ra &=&
-(u'v +v'u)\vec{\sigma}\times\hat{r}-i(u'v-v'u)\hat{r},   \non \\
\la{q'}|\vec{A}| {q}\ra
&=&\,(u'u-v'v)\vec{\sigma}+2v'v\,\hat{r}\vec{\sigma}\cdot\hat{r}.
 \en
The four-quark operators $O_1=(\bar{c}b)(\bar{d}u)$ and
$O_2=(\bar{c}u)(\bar{d}b)$  can be written as $O_1(x)=6(\bar
{c}b)_1(\bar{d}u)_2$ and $O_2=6(\bar{c}u)_1(\bar{d}b)_2$, where
the subscript $i$ on the r.h.s. of $O_{1,2}$ indicates that the
quark operator acts only on the $i$th quark in the baryon wave
function. It follows from Eq. (\ref{m.e.}) that the PC matrix
elements have the form
 \be \label{bagint}
 \int r^2dr\la q_1'q_2'|(\bar cb)_1(\bar
 du)_2|q_1q_2\ra &=& (-X_1+X_2)-{1\over
 3}(X_1+3X_2)\vec{\sigma}_1\cdot\vec{\sigma}_2,  \non \\
 \int r^2dr\la q_1'q_2'|(\bar db)_1(\bar
 cu)_2|q_1q_2\ra &=& (X_1+X_2)-{1\over
 3}(-X_1+3X_2)\vec{\sigma}_1\cdot\vec{\sigma}_2,
 \en
where the bag integrals are defined in Eq. (\ref{bagX}) and use
has been made of
  \be
  \int d\Omega\,\hat{r}_i\hat{r}_j={\delta_{ij}\over 3}\int d\Omega,
  \en
and those terms odd in $\hat{r}$ have been dropped since they
vanish after spatial integration. Note that we have applied the
isospin symmetry on the quark wave functions, namely
$u_d(r)=u_u(r)$ and $v_d(r)=v_u(r)$, to derive Eq. (\ref{bagint}).

We also need the spin-flavor wave functions of the baryons involved
such as
 \be \label{spin-flavor}
\Lambda_c^+ \up&=& {1\over\sqrt{6}}[(cud-cdu)\chi_A+(12)+(13)], \non \\
\Sigma_b^+ \up&=& {1\over\sqrt{3}}[buu\chi_s+(12)+(13)], \\
p^+ \up&=& {1\over\sqrt{3}}[duu\chi_s+(12)+(13)], \non
 \en
where $abc\chi_s=(2a^\dw b^\up c^\up-a^\up b^\up c^\dw-a^\up b^\dw
c^\up)/ \sqrt{6}$, $abc\chi_A=(a^\up b^\up c^\dw-a^\up b^\dw
c^\up)/\sqrt{2}$, and $(ij)$ means permutation for the quark in
place $i$ with the quark in place $j$. Applying Eq.
(\ref{spin-flavor}) yields
 \be \label{baryonm.e.}
 \la \Lambda_c^+\up|b_{1c}^\dagger b_{1b}b_{2d}^\dagger
 b_{2u}|\Sigma_b^+\up\ra &=& 0, \non \\
 \la \Lambda_c^+\up|b_{1c}^\dagger b_{1b}b_{2d}^\dagger
 b_{2u}(\vec{\sigma}_1\cdot\vec{\sigma}_2)|\Sigma_b^+\up\ra &=& {1\over \sqrt{6}}, \non \\
  \la \Lambda_c^+\up|b_{1c}^\dagger b_{1u}b_{2d}^\dagger
 b_{2b}|\Sigma_b^+\up\ra &=& {1\over 2\sqrt{6}}, \\
  \la \Lambda_c^+\up|b_{1c}^\dagger b_{1u}b_{2d}^\dagger
 b_{2b}(\vec{\sigma}_1\cdot\vec{\sigma}_2)|\Sigma_b^+\up\ra &=& -{1\over 2\sqrt{6}}. \non
 \en
In the above equation $b^\dagger_{1q'}~(b_{1q})$ denotes a quark
creation (destruction) operator acting on the first quark in the
baryon wave function. It is easily seen that $\la
\Lambda_c^+|O_1|\Sigma_b^+\ra=-\la \Lambda_c^+|O_2|\Sigma_b^+\ra$
and hence $\la \Lambda_c^+|O_+|\Sigma_b^+\ra=0$, as it should be.
The PC matrix element in Eq. (\ref{PCm.e.})
 \be
 \la \Lambda_c^+|O_-^{\rm PC}|\Sigma_b^+\ra=-{4\over\sqrt{6}}(X_1+3X_2)(4\pi)
 \en
then follows from (\ref{bagint}) and (\ref{baryonm.e.}).

For numerical estimates of the bag integrals $X_1$ and $X_2$, we
use the bag parameters
 \be
&& m_u=m_d=0,~~~m_s=0.279\,{\rm GeV},~~~m_c=1.551\,{\rm
GeV},~~~m_b=5.0\,{\rm GeV},  \non \\    &&
x_u=2.043\,,~~~x_s=2.488\,,~~~x_c=2.948\,,~~~x_b=3.079\,,~~~
R=5.0\,{\rm GeV}^{-1}.
 \en

%%%%%%%%%%%%%%%%%%%%%%%%%%%%%%%%%%%%%%%%%%%%%%%%%%%%%%%%

\end{document}